\documentclass[twocolumn,aps,superscriptaddress,nofootinbib,floatfix,linenumbers]{revtex4}
\usepackage{epsfig,bm,feynmf}
\usepackage{graphics}
\usepackage{amsmath}
\usepackage{mathrsfs}
\usepackage{appendix}
\usepackage{bm}
\usepackage[normalem]{ulem}  
\usepackage[dvips]{color} 

\renewcommand{\sout}{\bgroup \color{red} \ULdepth=-.5ex \ULset}

\begin{document}
\title{Impact of Glasma on heavy quark observables in nucleus-nucleus collisions at LHC}

\author{Yifeng Sun}
\email{sunyfphy@lns.infn.it}
\affiliation{Laboratori Nazionali del Sud, INFN-LNS, Via S. Sofia 62, I-95123 Catania, Italy}

\author{Gabriele Coci}
\email{coci@lns.infn.it}
\affiliation{Laboratori Nazionali del Sud, INFN-LNS, Via S. Sofia 62, I-95123 Catania, Italy}
\affiliation{Centro Siciliano di Fisica Nucleare e Struttura della Materia, CSFNSM, Via S. Sofia 64, I-95125 Catania, Italy}
\affiliation{Department of Physics and Astronomy, University of Catania, Via S. Sofia 64, 1-95125 Catania, Italy}

\author{Santosh Kumar Das}
\email{santosh@iitgoa.ac.in}
\affiliation{School of Physical Science, Indian Institute of Technology Goa, Ponda-403401, Goa, India}

\author{Salvatore Plumari}
\email{salvatore.plumari@ct.infn.it}
\affiliation{Department of Physics and Astronomy, University of Catania, Via S. Sofia 64, 1-95125 Catania, Italy}
\affiliation{Laboratori Nazionali del Sud, INFN-LNS, Via S. Sofia 62, I-95123 Catania, Italy}

\author{Marco Ruggieri}
\email{ruggieri@lzu.edu.cn}
\affiliation{School of Nuclear Science and Technology, Lanzhou University, 222 South Tianshui Road, Lanzhou
730000, China}

\author{Vincenzo Greco}
\email{greco@lns.infn.it}
\affiliation{Laboratori Nazionali del Sud, INFN-LNS, Via S. Sofia 62, I-95123 Catania, Italy}
\affiliation{Department of Physics and Astronomy, University of Catania, Via S. Sofia 64, 1-95125 Catania, Italy}

\date{\today}

\begin{abstract}
In the pre-thermal equilibrium stage of relativistic heavy-ion collisions, a strong quasi-classical transverse gluon 
field emerges at about $\tau_0 \simeq 0.1 \, \rm fm/c$ and evolves together with their longitudinal counterparts
according to the classical Yang-Mills (CYM) equations. 
Recently it has been shown that these fields induce a diffusion of charm quarks  
in momentum space resulting in a tilt of their spectrum without a significant drag.
We find that in nucleus-nucleus collisions at LHC such a novel dynamics of charm quarks leads to an initial enhancement of the nuclear modification factor ($R_{AA}$)  at $p_T$ larger than 2 GeV$/c$ contrary to the standard lore. 
Moreover, the same dynamics leads to a  larger final elliptic flow ($v_2$)  inducing a relation between $R_{AA}$ and $v_2$ that is quite
close to the experimental measurements. 
Our study also shows that such an initial pre-thermal stage is unlikely to be described in terms of  
a standard drag and diffusion dynamics,
because even if one tune such coefficients to reproduce the same $R_{AA}(p_T)$ this 
would imply a significantly smaller $v_2$.

\end{abstract}
\keywords{evolving glasma, heavy quark diffusion, nuclear modification factor, elliptic flow}

\maketitle

\section{Introduction}
The relativistic heavy-ion program provides the possibility to scrutiny Quantum Chromodynamics (QCD) phase diagram at
finite temperature and density in the region where a transition of nuclear matter into a plasma
of quarks and gluons (QGP) has been predicted.
In particular, in the last decade  the main properties of such a matter have been studied
both at the Relativistic Heavy Ion Collider (RHIC) and at the Large Hadron Collider (LHC), 
clarifying that the high temperature
medium created in high energy nuclear collisions is characterized by a large scattering rate
and thus a low shear viscosity over entropy density ratio, in disagreement with the naive
expectation of a weakly coupled plasma.

Heavy quarks
(Charm and Beauty) have been considered to have a unique role in such a study since they are generated in the early stage
$\tau_0 \simeq m_{HQ}^{-1} < O(10^{-1})\, \rm fm/c$ according to next-to leading order perturbative QCD and 
hence are witness of the entire evolution of the QGP; 
furthermore because $M_{HQ} >> T_c$  they still preserve their ‘‘identity’’  
at hadronization by picking-up a light quark or undergoing an independent fragmentation.
These properties, together with a thermalization time $\tau_{therm}$ that is comparable to the lifetime of the
QGP phase~\cite{Dong:2019unq,Greco:2017rro}, makes them a probe able to preserve key information about the time evolution of their interaction in the 
hot QCD medium.  In addition, it is in perspective possible to perform a direct comparison of the transport properties
of the heavy quarks with lattice QCD (lQCD) calculations. Recently, it has been shown
that the determination of the space-diffusion transport coefficient $D_s$ from the phenomenology \cite{Dong:2019unq,Rapp:2018qla} 
is in agreement with first lQCD calculations in quenched approximation \cite{Banerjee:2011ra,Kaczmarek:2014jga}, even if within still significant uncertainty 
both in the phenomenology as well as in the lQCD approach.

The study of the HQ physics in $AA$ collisions has been successful and also has clearly shown that
in the low momentum regime $p_T < 10 \,\rm GeV$ the interaction is strongly non-perturbative and implies a 
space diffusion coefficient $2\pi T\, D_s \sim 2-5$ around $T_c$~\cite{Dong:2019unq}.
 The determination of such a coefficient is mainly driven by the phemonemological prediction of two main observables:
the nuclear modification factor $R_{AA}$ and the elliptic flow $v_2$. However, there has been always a 
tension between these two observables that are hard to be correctly predicted simultaneously~\cite{Cao:2016gvr,Xu:2018gux,Dong:2019unq,vanHees:2005wb,Das:2010tj,Plumari:2011mk,Cao:2018ews}.
A significant part of such a tension is reduced when an increasing temperature dependence of $D_s$
 and an hadronization by coalescence plus fragmentation  are included, as discussed in \cite{Das:2015ana}.
However, especially at the LHC energy such a tension persists and is currently partially moderated by the still significant
large error bars in the experimental data, especially for the elliptic flow \cite{Scardina:2017ipo,Rapp:2018qla}. 

We notice that while the dynamics of the HQs in the QGP phase has been thoroughly studied and also the 
possible impact of the later stage of hadronic re-scattering has been discussed in several approaches
\cite{Nahrgang:2016lst,Song:2015ykw,Cao:2014fna,He:2012df,Das:2016llg},
the dynamics of HQs in the early stage has been addressed only recently~\cite{Ruggieri:2018rzi}.
The early stage dynamics can be quite relevant especially for HQ's considering their short formation time and the
fact that HQ thermalization time is comparable to the lifetime of the QGP. This, in fact,  implies a larger sensitivity  to the early time 
evolution of the HQ during the fireball expansion, hence potentially keeping memory of the initial dynamics.
Indeed, recently, it has been shown that HQs dynamics should be particularly sensitive to both the initial
electromagnetic field and the tilted initial condition of the medium \cite{Das:2016cwd, Chatterjee:2017ahy}, 
a prediction that appears to be confirmed by early
experimental results at both RHIC and LHC energies \cite{Singha:2018cdj}.

According to the the color glass 
 condensate (CGC) effective theory~\cite{McLerran:1993ni,McLerran:1993ka,McLerran:1994vd,Iancu:2000hn}, 
the dense gluon system produced by the interaction of two colored glasses immediately after the collision 
can be described in terms of classical longitudinal fields named the Glasma,
whose evolution in the early stage can be described by means of the classical Yang-Mills (CYM) equations.
Recent works have investigated the impact of the propagation  of charm and beauty 
in the Glasma\cite{Mrowczynski:2017kso,Ruggieri:2018rzi,Ruggieri:2018ies,Song:2015jmn} with particular reference to
the diffusion in momentum space.
Here we study for the first time how the interaction of the heavy quarks with the Glasma affects the dynamics of HQs
in AA collisions; we achieve this by implementing a simulation of the HQs dynamics
including both the glasma and the QGP stages. In this preliminary study we have not implemented the longitudinal expansion in the YM evolution: in order to overcome this problem we have used a small value of the saturation scale, in agreement with the lowest bound for the estimate of this scale for AA collisions at the LHC energy~\cite{Ruggieri:2018rzi}.
We show  that as a result of the diffusion of HQs in the Glasma, 
there is a significant modification of the relation between $R_{AA}$
and $v_2$; we expect a significant impact also on several other observables, both in $pA$ and $AA$ collisions.
Although a systematic study is feasible in perspective, here we report specifically on Pb-Pb collisions at $\sqrt{s_{NN}}=5.02$ TeV
in order to emphasize the novel idea.

In this Letter, we firstly discuss how we set up the initial condition and the evolution of HQs in the Glasma
by means of the Wong equations~\cite{Wong:1970fu}; then we discuss how this initial stage evolution
is embedded in a modified Fokker-Planck equation that is able to reproduce a very similar dynamics, but 
in addition allows for the implementation of the subsequent standard HQs evolution in the QGP. 
We  then focus on the impact of the initial stage dynamics 
on $R_{AA}$ and $v_2$ estimating the uncertainty on these observables due to the
lifetime of the Glasma stage; in addition to this, 
we also show the difference with standard approaches that neglect the diffusion of HQs in the glasma phase.
Finally we draw our conclusions with an outlook for upcoming studies.

\section{Model setup}

Within the CGC effective theory the pre-thermal equilibrium stage of the Pb-Pb collisions can be described 
in terms of strong gluon fields, namely the Glasma,
that evolves according to the CYM equations. 
In the temporal gauge $A_{0}^a=0$, the Hamiltonian density takes the following form~\cite{Kunihiro:2010tg,Iida:2014wea}
\begin{equation}
H = \frac{1}{2}\sum_{a,i}E_i^a(x)^2 + \frac{1}{4}\sum_{a,i,j}F_{ij}^a(x)^2,
\label{eq:H} 
\end{equation}
where 
\begin{equation}
F_{ij}^a(x) = \partial_i A_j^a(x) - \partial_j A_i^a(x)  + \sum_{b,c}f^{abc} A_i^b(x) A_j^c(x).
\label{eq:Fij}
\end{equation}
As in \cite{Ruggieri:2018rzi,Ruggieri:2018ies}  we deal with the SU(2) gauge theory for simplicity, 
we thus have  $f^{abc} = \varepsilon^{abc}$ with $\varepsilon^{123} = +1$.
The CYM equations are 
\begin{eqnarray}
\frac{dA_i^a(x)}{dt} &=& E_i^a(x),\\
\frac{dE_i^a(x)}{dt} &=& \sum_j \partial_j F_{ji}^a(x) + 
\sum_{b,c,j} f^{abc} A_j^b(x)  F_{ji}^c(x) \label{eq:CYM_el},
\end{eqnarray}
which we solve in a static box in three spatial 
dimensions~\cite{Kunihiro:2010tg,Iida:2014wea,Ruggieri:2017ioa,Ruggieri:2018rzi}. We have adopted periodic boundary conditions at finite time. We neglect the longitudinal expansion in the YM stage, so the natural coordinate system to use is the ($t,x,y,z$) with $t$ the lab time, ($x,y$) the transverse plane and $z$ the longitudinal coordinate; invariance along the longitudinal direction is assumed, analogously to the rapidity invariance of the expanding system. The box size is 4 fm$\times$4 fm in the transverse plane with a lattice spacing $a$=0.04 fm. 

The initialization of the classical gluon field is given by the standard Mclerran-Venugopalan (MV) 
model~\cite{McLerran:1993ni,McLerran:1993ka,McLerran:1994vd}, in which the transverse color charge densities $\rho_{a}$ on 
the nucleus A (same for B in an AB collision at high energy) 
are assumed to be distributed with zero mean and variance specified by
\begin{equation}
\langle \rho^a_A(\bm x_T)\rho^b_A(\bm y_T)\rangle = 
(g^2\mu_A)^2 \delta^{ab}\delta^{(2)}(\bm x_T-\bm y_T),
\label{eq:dfg}
\end{equation}
with $a,b=1,2,3$ for SU(2). We set $g^2 \mu_A = 3$ GeV in this study, in agreement with the estimate
of the saturation scale obtained within the IP-Sat model for the relevant LHC 
energy~\cite{Lappi:2007ku,Kowalski:2007rw,Ruggieri:2018rzi}. In order to determine these fields
we  firstly solve the Poisson equations for the gauge potentials
generated by the color charge distributions of the nuclei $A$ and $B$, namely
\begin{equation}
-\partial_\perp^2 \Lambda^{(A)}(\bm x_T) = \rho^{(A)}(\bm x_T)
\end{equation}
(a similar equation holds for the distribution belonging to $B$). Wilson lines are computed as
$
V^\dagger(\bm x_T) = e^{i \Lambda^{(A)}(\bm x_T)}$, 
$W^\dagger(\bm x_T) = e^{i \Lambda^{(B)}(\bm x_T)}$,
and the pure gauge fields of the two colliding nuclei are given by
$
\alpha_i^{(A)} = i V \partial_i V^\dagger$,
$\alpha_i^{(B)} = i W \partial_i W^\dagger$.
In terms of these fields the solution of the CYM in the forward light cone
at initial time, namely the Glasma gauge potential, 
can be written as 
$A_i = \alpha_i^{(A)} + \alpha_i^{(B)}$~ for $i=x,y$ and $A_z = 0$,
and the initial longitudinal Glasma fields are
\begin{eqnarray}
&& E^z = i\sum_{i=x,y}\left[\alpha_i^{(B)},\alpha_i^{(A)}\right], \label{eq:f1}\\
&& B^z = i\left(
\left[\alpha_x^{(B)},\alpha_y^{(A)}\right]  + \left[\alpha_x^{(A)},\alpha_y^{(B)}\right]  
\right),\label{eq:f2}
\end{eqnarray}
while the transverse fields are vanishing. It is useful to remind that the initial color charges give no contribution to the gluon fields at finite time. This is the standard assumption of any calculation based on the effective theory of the color-glass-condensate. These details have been discussed many times in the literature, see for example~\cite{Lappi:2006fp}.


Charm and anti-charm quarks are produced in the pre-thermal equilibrium stage of relativistic heavy ion collisions; 
the formation time is $\tau_\mathrm{form} \simeq m_{c}^{-1} \approx 0.1$ fm$/c$. 
The equations of motion of these heavy color probes in the pre-equilibrium stage are the Wong equations~\cite{Wong:1970fu}
\begin{eqnarray}
&&\frac{d x_i}{dt} = \frac{p_i}{E},\\
&&E\frac{d p_i}{dt} = Q_a F^a_{i\nu}p^\nu,
\label{wong1}
\end{eqnarray}
with $i=x,y,z$ and $E=\sqrt{\bm p^2 + m^2}$. The conservation of the color charge is guaranteed by the additional equation
\begin{eqnarray}
&&E\frac{d Q_a}{dt} = - Q_c\varepsilon^{cba} \bm A_b\cdot\bm p,
\label{wong2}
\end{eqnarray}
with the color charge $Q_a$ ($a=$1,2,3) of charm quarks  
that we initialize randomly in the range $(-1,+1)$ with uniform distribution for each color charge.

We study the Pb+Pb collisions at LHC $\sqrt{s_{NN}}=5.02$ TeV at fixed impact parameter $b=9.25$ fm in 
order to simulate the collisions at centrality range $30\%-50\%$. 
At the initial time $\tau_{form}=0.1$ fm$/c$, the position coordinate of charm quarks are distributed 
according to the binary nucleon-nucleon collisions profile derived from the standard Glauber model, while 
the initial momentum of the charm quarks are distributed from the one obtained within 
Fixed Order+Next-to-Leading Log(FONLL) QCD, which we parametrize it as:
\begin{eqnarray}
&&\frac{dN}{d^2p_T}=\frac{x_0}{(1+x_3 {p_T}^{x_1})^{x_2}},
\end{eqnarray}
where the parameters are $x_0=20.28$, $x_1=1.950$, $x_2=3.136$ and $x_3=0.0751$ respectively. HQs are distributed with $p_z$=0 at the initial time, because in our work we are considering the rapidity range around $y$=0. Starting with this initial conditions for the charm distribution and with a background of evolving Glasma
as described above, we 
follow the evolution of the charm quarks by means of the Wong's equation, Eqs. \eqref{wong1} and \eqref{wong2}
until the formation of QGP at $\tau_{glasma}=0.3-0.5 \rm fm/c$ (which mean a glasma lifetime
in the range $0.2-0.4\, \rm fm/c$), where the label "glasma" stands for the time at which this
phase ends and the standard hydrodynamical QGP evolution starts.
We have considered such a range of values that represents a typical range for the beginning of the
hydrodynamical expansion of the QGP phase at LHC energies.

\begin{figure}[h]
\centering
\includegraphics[width=1.2\linewidth]{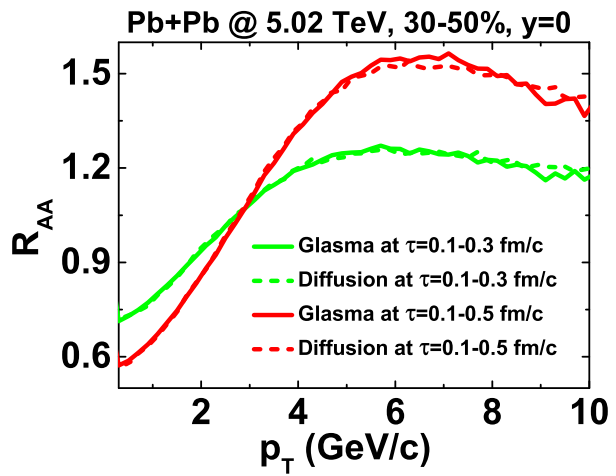}
\caption{(Color online) $R_{AA}$ of charm quarks affected by the glasma or diffusion dynamics for different evolving times for Pb+Pb 30\%-50\% centrality collisions at 5.02 TeV.}
\label{fig:cRAA}
\end{figure}

In Fig. \ref{fig:cRAA}, we show the results for the $R_{AA}(p_T)$ when we assume that the glasma
phase ends at $\tau_{glasma}=0.3\,\rm fm/c$ and at $\tau_{glasma}=0.5\, \rm fm/c$. The results shown have been obtained employing 75 configurations of the glasma fields. The glasma is seen to push charm quarks from low to high $p_T$, which is shown by the solid green and solid red 
lines of Fig.~\ref{fig:cRAA} and the $R_{AA}$ increases with $p_T$ up to 1.2 for $\tau_{glasma}=0.3\,\rm fm/c$ 
and 1.5 for $\tau_{glasma}=0.5\, \rm fm/c$.
Therefore, we find a significant enhancement 
of the $R_{AA}(p_T)$ at intermediate $p_T$ associated to a depletion at low momenta.
Such a dynamics is the result of the diffusion of
the heavy quarks in the evolving gluon fields
as already studied in the context of $pA$ collisions in \cite{Ruggieri:2018ies}.
Certainly in $AA$ collisions we have never observed such a behavior of $R_{AA}(p_T)$ because the initial stage
is followed by a long phase of in-medium charm quark scattering. The aim of the present work is however to
point out the impact on final observable of this initial glasma dynamics.
To this end it is necessary to merge this initial glasma phase
with the subsequent standard evolution of charm quarks in the QGP. 

The standard method applied for heavy quark (HQ) dynamics in QGP phase is to follow their evolution by means of Fokker-Planck equation, which is usually solved stochastically by the Langevin equations:
\begin{eqnarray}
&&d x_i = \frac{p_i}{E}dt,\\
&&{d p_i} = -\Gamma p_idt+C_{ij}\rho_j\sqrt{dt},
\end{eqnarray}
where $dx_i$ and $dp_i$ are the change of the coordinate and momentum of each HQ in each time step $dt$. $\Gamma$ and $C_{ij}$ are the drag and the covariance matrix 
related to the diffusion tensor by
\begin{eqnarray}
&&C_{ij}=\sqrt{2B_0}P_{ij}^{\perp}+\sqrt{2B_1}P_{ij}^{\parallel },
\label{tensor}
\end{eqnarray}
where $P_{ij}^{\perp}=\delta_{ij}-\frac{p_ip_j}{p^2}$ and $P_{ij}^{\parallel }=\frac{p_ip_j}{p^2}$ are the transverse and longitudinal projector operators respectively. We employ the standard assumption $B_0=B_1=D$ for all momenta in this study as Refs.~\cite{Moore:2004tg,vanHees:2005wb,Cao:2011et,vanHees:2007me,Das:2010tj}, though this is strictly valid only for $p \rightarrow 0$. Under this assumption, Eq. (\ref{tensor}) becomes simply $C_{ij}=\sqrt{2D}\delta_{ij}$.
This is the approach that most of the groups have used to study the heavy quark dynamics
 in $AA$ collisions \cite{Gossiaux:2008jv,Gossiaux:2009mk,Cao:2015hia,Xu:2017obm,vanHees:2005wb,vanHees:2007me,He:2011qa,Alberico:2011zy,Alberico:2013bza}.

At present an approach that starting from the colored glasma field evolves into the locally thermalized
quark-gluon plasma is missing. 
However we notice that the effect of the glasma fields resembles that of an anomalous Fokker-Planck diffusion 
without a sizeable drag that degrades the charm momentum, as can be seen in Fig.3 of Ref.\cite{Ruggieri:2018rzi}.
Because of this, we can simulate the dynamics of HQs in the pre-thermal equilibrium phase of heavy ion collisions by using the Fokker-Planck equation. To this end we modified the initial Langevin dynamics to mimic the glasma effect on charm quark transport
by gauging a momentum dependent diffusion coefficient to generate the same $R_{AA}(p_T)$ obtained within the
Wong equations dynamics.
The results are shown in Fig. \ref{fig:cRAA} by dashed lines for the two cases considered:
a glasma dynamics up to $\tau_{glasma}=0.3 \, \rm fm/c$ (green dashed line) and 
$\tau_{glasma}=0.5 \, \rm fm/c$ (red dashed line) and can be compared to the corresponding results from the Wong equations
shown by solid lines.
The results with the Langevin simulation are obtained with $\Gamma=0$ and a diffusion coefficient 
parametrized as $D(p)=D_0(1+cp)^2$, where $c$ affects only $R_{AA}$ at large $p_T$ regulating its increase 
by increasing the parameter $c$. 
To reproduce the $R_{AA}(p_T)$ due to the effect of the glasma, solid lines in Fig.\ref{fig:cRAA}, we find that 
$D_0=3.6$ GeV$^2/$fm, and $c=0.02$ GeV$^{-1}$ for the case $\tau_{glasma}=0.3 \, \rm fm/c$ 
and a slightly increased $c=0.025\, \rm GeV^{-1}$ to have a nearly perfect matching with the results
with Wong's equation also for  $\tau_{glasma}=0.5 \, \rm fm/c$ . 

Given that we can gauge the Fokker-Planck equation to mimic the initial stage dynamics up to $\tau_{Glasma}$
by mean of $D(p)$ as determined above,
we continue the charm dynamical evolution at $\tau > \tau_{Glasma}$ accordin to the standard charm
evolution given by the Fokker-Planck evolution with a drag $\Gamma$ and a diffusion $D$ related by the
fluctuation-dissipation theorem. We note that the diffusion coefficient $D(p)$ in the pre-thermal equilibrium glasma stage is larger than the one in thermally equilibrated QGP stage, which means that there exists a transition between these two stages. As there is a lack of model that describes this smooth transition, we thus simply switch the diffusion coefficient to the one in QGP phase at $\tau > \tau_{Glasma}$. The background medium is given by
the expanding bulk according to viscous hydrodynamics by a relativistic transport Boltzmann solved at
fixed shear viscosity to entropy density $\eta/s$ \cite{Ruggieri:2013ova,Plumari:2015cfa,Plumari:2019gwq}.
More specifically, we determine the heavy flavor in medium scattering employing the drag $\Gamma$ and the 
diffusion $D$ derived from  a quasi-particle model (QPM)~\cite{Das:2012ck,Berrehrah:2013mua,Berrehrah:2014kba}. The QPM approach accounts for the non-perturbative dynamics by $T-$dependent quasi-particle masses, with $m_q^2=1/3g^2(T)T^2$ and $m_g^2=3/4g^2(T)T^2$, plus a $T-$dependent background field known as a bag constant. 
with $g(T)$ tuned to fit the thermodynamics of the lattice QCD~\cite{Borsanyi:2010cj,Plumari:2011mk}.

This approach has been shown to lead to a good description of the experimental data for the $R_{AA}(p_T)$, 
both at RHIC and LHC employing  an enhancement factor $K \sim 2$ of the drag and diffusion coefficient.
This and similar approaches have allowed  first estimates of the space-diffusion coefficient, as discussed in \cite{Scardina:2017ipo,Prino:2016cni,Dong:2019unq}.
However all these approaches do not consider an evolving Glasma stage and have an initialization time 
$\tau_0 \simeq 0.3-0.6 \, \rm fm/c$.
 

\begin{figure}
\centering
\includegraphics[width=1\linewidth]{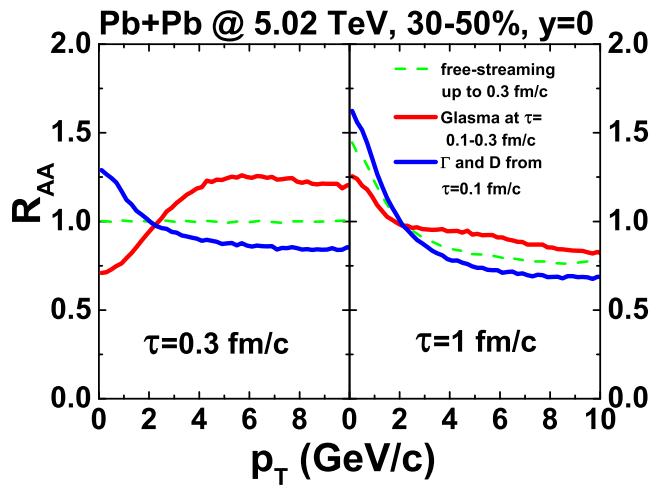}
\caption{(Color online) $R_{AA}$ of charm quarks for cases:
(a) free streaming up to $0.3\, \rm fm/c$ followed by drag and diffusion in a hydro bulk (red solid line); 
(b) glasma dynamics up to $0.3\, \rm fm/c$ and then evolution as in (a) shown by green dashed line; 
(c) drag and diffusion dynamics starting at $\tau_0= 0.1 \,\rm fm/c$ shown by blue solid line. See text for more details.}
\label{fig:tcRAA}
\end{figure}

\begin{figure}[h]
\centering
\includegraphics[width=1\linewidth]{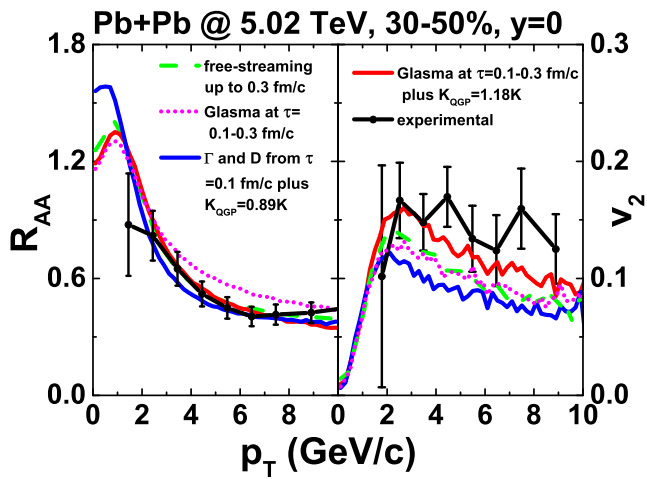}
\caption{(Color online) 
$R_{AA}$ (left) and $v_2$ (right) of D mesons for the cases:
(a) free streaming up to $0.3\, \rm fm/c$ followed by drag and diffusion in a hydro bulk (green dashed line); 
(b) glasma dynamics up to $0.3\, \rm fm/c$ and then evolution as in (a) shown by magenta dotted line; 
(c) drag and diffusion dynamics starting at $\tau_0= 0.1 \,\rm fm/c$ shown by blue solid line with K factor decreasing by 11\% relative to (a); (d) glasma dynamics up to $0.3\, \rm fm/c$ and then evolution with K factor increasing by 18\% relative to (a) shown by red solid line. Experimental data are taken from Refs.~\cite{Barbano:2017bcu,Acharya:2017qps}.}
\label{fig:DRAA}
\end{figure}

We choose three cases in order to see the effect of the glasma field: 
a) free streaming up to $\tau= 0.3 \rm \,fm/c$ followed by a Langevin dynamics in a hydro-like bulk for $\tau>0.3 \, \rm fm/c$; 
b) only diffusion mimicking the glasma evolution up to $\tau_{glasma}=0.3 \,\rm fm/c$ followed by standard Langevin dynamics 
with drag and diffusion in a hydro-like but at $\tau>0.3\, \rm fm/c$;
c) drag and diffusion from QPM  in a Langevin dynamics starting at $\tau=0.1$ fm$/c$. 
This third case is included to see what happens if one simulate the initial stage just starting with an initial time $\tau_0=0.1 \, \rm fm/c$
letting charm quarks to undergo scattering in the QGP medium.
In all these case we have tuned by a K factor the $\Gamma$ and $D$ in the QGP phase to reproduce a very similar final
$R_{AA}(p_T)$ that reasonably describes the
experimental data, as can be seen from Fig. \ref{fig:DRAA} (right panel).

The aim of this Letter is to point out that heavy-quark interaction with glasma fields induces a dynamics that is opposite to the 
standard heavy-quark in medium scattering with the bulk medium. We can see in
 Fig.~\ref{fig:tcRAA} the transverse momentum dependence of $R_{AA}$ of charm quarks at time $\tau=0.3$ (left panel) and 1 fm$/c$ (right panel). 
In the left panel of Fig.~\ref{fig:tcRAA}, we can see that at time $ 0.3\, \rm  fm/c$, the glasma leads to the enhancement of $R_{AA}$ at high $p_T$ (red solid line), while charm propagation in a hydro bulk produces a significant suppression 
at $p_T>2 \, \rm GeV$ shown
by the blue solid line. The right panel of Fig.~\ref{fig:tcRAA} shows that drag and diffusion leads to the suppression of $R_{AA}$ of high $p_T$ in an efficient way, which can be seen by the 
significant dropping of the differences compared to the left panel
between the three cases  already at $\tau \simeq 1 \, \rm fm/c$.
From this we can already understand that to consider the evolution under the gluonic fields at early time is quite different
from moving the initial time evolution of the QGP phase down to $\tau_0=0.1 \, \rm fm/c$.
After the evolution of charm quarks for the three different cases mentioned in the above section, we hadronize charm quarks 
to $D$ mesons by a standard hadronization by fragmentation as done in \cite{Scardina:2017ipo,vanHees:2007me}. 

In the left panel of Fig.~\ref{fig:DRAA}, we show $R_{AA}$ for the three cases, and compare them to the experimental results for the same collision system and centrality~\cite{Barbano:2017bcu,Acharya:2017qps}. 
It is seen that we can nicely reproduce almost the same $R_{AA}$ in all three cases, but the associated 
$v_2$ is quite different (right panel). It should be noted by magenta dotted line that, if the system evolves after the glasma phase at $\tau=0.3$ fm$/c$ with K factor same as the case without the glasma phase, $v_2$ does not change so much while $R_{AA}$ will increase.

In Fig. \ref{fig:DRAA}, the dashed green line shows $v_2$ of standard evolution in a hydrodynamic bulk
starting at $\tau_0= 0.3 \, \rm fm/c$, which determines values of $v_2$ below the experimental results. However, with the inclusion of the Glasma 
field before the formation of QGP, $v_2$ is significantly larger,  about a $15-20\%$
at $p_T$ larger than 2 $\rm GeV/c$ and is in the lower limit of experimental results, which is shown by the solid red line 
of right panel of Fig.~\ref{fig:DRAA}. If we start the hydrodynamic evolution earlier at $\tau_0=0.1 \, \rm fm/c$, 
which is shown by the solid blue line, $v_2$ decreases further by about a $20\%$ at $p_T>2$ GeV$/c$. 
This shows that the initial glasma dynamics cannot be simply simulated by decreasing the intial time $\tau_0$ 
as done for example in \cite{Das:2015aga}. In fact this last case would lead to estimate about a $30\%$ smaller 
ellitptic flow of the D mesons which
also means a $v_2(p_T)$ quite smaller with respect to the experimental data.

The generation of a larger $v_2$, when a glasma phase is taken into account, is related with the initial enhancement
of $R_{AA}(p_T)$. In fact as discussed in Ref. \cite{Das:2015ana} $R_{AA}$ and $v_2$ are correlated in such a way
that a lower nuclear modification factor is associated to a larger $v_2$. However
the size of this anti-correlation depends on the time evolution of $R_{AA}$. The initial enhancement due to the 
diffusion of charm quarks in the evolving Glasma implies that during the QGP phase more interaction 
is needed in order to get a $R_{AA}(p_T)$ that agrees with the experimental data. 
This means that charm quarks interact more with the medium
when the bulk has a larger ellipticity, so that the bulk itself can transfer it to the charm quark more efficiently. 

\begin{figure}[h]
\centering
\includegraphics[width=1\linewidth]{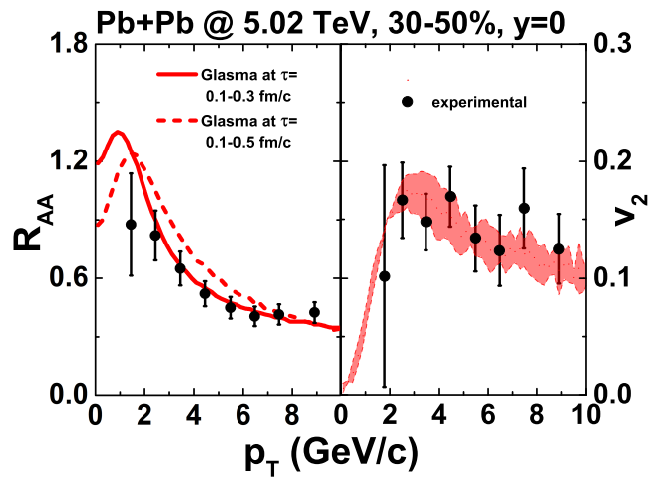}
\caption{(Color online) 
$R_{AA}(p_T)$ (left) and uncertainty band on $v_2(p_T)$ (right) of D mesons associated to a glasma dynamics
in the range $0.2-0.4\, \rm fm/c$ followed by standard Fokker-Planck evolution in the QGP. Experimental data are taken from Refs.~\cite{Barbano:2017bcu,Acharya:2017qps}.}
\label{fig:DRAAv2compare}
\end{figure}

As the lifetime of the glasma field is not yet clearly defined, we explore also the 
impact of its uncertainty varying it in the range $\tau_{glasma}=0.3-0.5 \, \rm fm/c$ that is the
usual time interval within which a hydrodynamical expansion of the bulk QGP matter is considered.
In Fig.\ref{fig:DRAAv2compare} (right panel) the band indicates the uncertainty in $v_2(p_T)$ associated to
the uncertainty in the glasma lifetime that appears comparable with the current error bars of the 
data with the upper bound associated to the
largest lifetime considered. It appears a quite good agreement with both the $R_{AA}$  and $v_2$ of the experimental data.
We also emphasize that usually to have a reasonable agreement with both the  $R_{AA}$  and $v_2$ 
in the phenomenological model one prefers to chooce an initial time $\tau_0$ for the evolution
in the range $0.3-0.6 \rm fm/c$ with a preference on larger times because otherwise the
associated $v_2$ is smaller, in fact this is shown by the blue solid line in Fig. \ref{fig:DRAA} (left panel).
Indeed assuming that nothing relevant happens in the HQ dynamical evolution,
is not in general really justified considered also the short HQ formation time $\tau_{form} \leq O(10^{-1}) \, \rm fm/c$.
The mechanism of the initial $R_{AA}$ enhancement allows to not
discard the initial phase while acquiring an even better description of the experimental observables.

\section{Conclusions and Discussions}
In this study we have pointed out the effect of the initial gluon field on the evolution of charm quarks in high energy Pb+Pb 
collisions at 5.02 TeV. We have found that the gluon field can push the 
charm quarks from low to high $p_T$, and this effect can be seen as
an initial diffusion of charm quarks in momentum space without a degrading drag force. 
The effect discussed here is qualitatively different w.r.t. the standard dynamics of HQ in the hot QCD medium, 
because it implies a rise and fall of the $R_{AA}(p_T)$ during its time evolution. 
Using the Fokker-Planck equation extended to include both the effect of the glasma and QGP, we 
have found that  the correlation between the $R_{AA}$ and $v_2$ of $D$ mesons is modified leading to
a final large $v_2$ at the same $R_{AA}$ when
the evolution in the glasma is considered. This points towards a better description of both observables and
we have discussed that this can be understood as an effect due to the delay in the formation of $R_{AA}$.
We have also found that a longer lifetime of the glasma  can lead to a larger effect on the enhancement of $v_2$,
but already for $\tau_{glasma} \sim 0.3 \, \rm fm/c$ the effect can be quite large, about a $20\%$.
We remark that the longitudinal expansion of the colliding systems in the glasma phase is not included in this study. 
This may affect quantitatively the effect on $R_{AA}$ and $v_2$, and should be considered in future studies. 
However, the qualitative result presented here will not be affected by the presence of the longitudinal expansion,
as this will have the effect to dilute the energy density but it will not affect the physical mechanisms described. We note that there is a discontinuity when we switch the diffusion coefficient from glasma phase to QGP phase. A complete modeling should be developed to describe the evolution of the glasma phase in a QGP one through a mix stage. This is certainly a challenge for future developments.
We did not include the cold nuclear matter effect which affect $R_{AA}$
especially at $p_T< 2\, \rm GeV$, even if it would enhance the nuclear
modification factor at intermediate $p_T$ further contributing to the main effect
discussed here. We have not included for simplicity the back-reaction on the HQs in the glasma phase. This has to be certainly studied in future developments. However, considering that the only energy scale is $g^2\mu$ which is in the range 3-5 GeV, we may expect that the effect of the gluon radiation will be not substantial at least up to this $p_T$ scale.

In this Letter we have presented the first results on the two main observables in HIC, $R_{AA}$ and $v_2$. However
a thorough understanding of the initial stage dynamics is a timely fundamental task and should affect at least two
other observables like the  triggered $D- \bar D$ angular correlation and the splitting in the directed flow $v_1$ 
that recently has been shown to probe the initial strong electromagnetic field that reaches its maximum value in the 
same time range of the glasma stage \cite{Das:2016cwd}.
In our opinion, studies in this direction have to be pursued because they
will also allow a unified description between the dynamics in $pA$ and in $AA$ collisions.
Furthermore such a direction joins the current lively activity in the study of the initial 
stage of ultra-relativistic collisions to the
HQ physics.

\section*{ACKNOWLEDGEMENTS}
The work of  Y.S. is supported by a INFN post-doc fellowship within the national SIM project; 
S.K.D. and M. R. are supported by the National Science Foundation of China (Grants No.11805087 and No. 11875153)
and by the Fundamental Research Funds for the Central Universities (grant number 862946).


\begin{thebibliography}{0}
\expandafter\ifx\csname natexlab\endcsname\relax\def\natexlab#1{#1}\fi
\expandafter\ifx\csname bibnamefont\endcsname\relax
  \def\bibnamefont#1{#1}\fi
\expandafter\ifx\csname bibfnamefont\endcsname\relax
  \def\bibfnamefont#1{#1}\fi
\expandafter\ifx\csname citenamefont\endcsname\relax
  \def\citenamefont#1{#1}\fi
\expandafter\ifx\csname url\endcsname\relax
  \def\url#1{\texttt{#1}}\fi
\expandafter\ifx\csname urlprefix\endcsname\relax\def\urlprefix{URL }\fi
\providecommand{\bibinfo}[2]{#2}
\providecommand{\eprint}[2][]{\url{#2}}

\end{thebibliography}


\begin{thebibliography}{00}
\bibitem{Dong:2019unq}
  X.~Dong and V.~Greco,
  Prog.\ Part.\ Nucl.\ Phys.\  {\bf 104} (2019) 97.
  
  \bibitem{Greco:2017rro}
  V.~Greco,
  Nucl.\ Phys.\ A {\bf 967} (2017) 200.
  
  \bibitem{Rapp:2018qla}
  R.~Rapp {\it et al.},
  Nucl.\ Phys.\ A {\bf 979} (2018) 21.
  
  \bibitem{Banerjee:2011ra}
  D.~Banerjee, S.~Datta, R.~Gavai and P.~Majumdar,
  Phys.\ Rev.\ D {\bf 85} (2012) 014510.
  
  \bibitem{Kaczmarek:2014jga}
  O.~Kaczmarek,
  Nucl.\ Phys.\ A {\bf 931} (2014) 633.
  
  
  \bibitem{Cao:2016gvr}
  S.~Cao, T.~Luo, G.~Y.~Qin and X.~N.~Wang,
  Phys.\ Rev.\ C {\bf 94} (2016) no.1,  014909.
  
  \bibitem{Xu:2018gux}
  Y.~Xu {\it et al.},
  Phys.\ Rev.\ C {\bf 99} (2019) no.1,  014902.
  
  \bibitem{vanHees:2005wb}
  H.~van Hees, V.~Greco and R.~Rapp,
  Phys.\ Rev.\ C {\bf 73} (2006) 034913.
  
  \bibitem{Das:2010tj}
  S.~K.~Das, J.~e.~Alam and P.~Mohanty,
  Phys.\ Rev.\ C {\bf 82} (2010) 014908.
  
  \bibitem{Plumari:2011mk}
  S.~Plumari, W.~M.~Alberico, V.~Greco and C.~Ratti,
  Phys.\ Rev.\ D {\bf 84} (2011) 094004.
  
  \bibitem{Cao:2018ews}
  S.~Cao {\it et al.},
  arXiv:1809.07894 [nucl-th].
  
  \bibitem{Das:2015ana}
  S.~K.~Das, F.~Scardina, S.~Plumari and V.~Greco,
  Phys.\ Lett.\ B {\bf 747} (2015) 260.
  
  \bibitem{Scardina:2017ipo}
  F.~Scardina, S.~K.~Das, V.~Minissale, S.~Plumari and V.~Greco,
  Phys.\ Rev.\ C {\bf 96} (2017) no.4,  044905.
  
  \bibitem{Nahrgang:2016lst}
  M.~Nahrgang, J.~Aichelin, P.~B.~Gossiaux and K.~Werner,
  Phys.\ Rev.\ C {\bf 93} (2016) no.4,  044909.
  
  \bibitem{Song:2015ykw}
  T.~Song, H.~Berrehrah, D.~Cabrera, W.~Cassing and E.~Bratkovskaya,
  Phys.\ Rev.\ C {\bf 93} (2016) no.3,  034906.
  
  \bibitem{Cao:2014fna}
  S.~Cao, Y.~Huang, G.~Y.~Qin and S.~A.~Bass,
  J.\ Phys.\ G {\bf 42} (2015) no.12,  125104.
  
  \bibitem{He:2012df}
  M.~He, R.~J.~Fries and R.~Rapp,
  Phys.\ Rev.\ Lett.\  {\bf 110} (2013) no.11,  112301.
  
  \bibitem{Das:2016llg}
  S.~K.~Das, J.~M.~Torres-Rincon, L.~Tolos, V.~Minissale, F.~Scardina and V.~Greco,
  Phys.\ Rev.\ D {\bf 94} (2016) no.11,  114039.
  
  \bibitem{Ruggieri:2018rzi}
  M.~Ruggieri and S.~K.~Das,
  Phys.\ Rev.\ D {\bf 98} (2018) no.9,  094024.
  
  \bibitem{Das:2016cwd}
  S.~K.~Das, S.~Plumari, S.~Chatterjee, J.~Alam, F.~Scardina and V.~Greco,
  Phys.\ Lett.\ B {\bf 768} (2017) 260.
  
  \bibitem{Chatterjee:2017ahy}
  S.~Chatterjee and P.~Bozek,
  Phys.\ Rev.\ Lett.\  {\bf 120} (2018) no.19,  192301.
  
  \bibitem{Singha:2018cdj}
  S.~Singha [STAR Collaboration],
  Nucl.\ Phys.\ A {\bf 982} (2019) 671.
  
  \bibitem{McLerran:1993ni}
  L.~D.~McLerran and R.~Venugopalan,
  Phys.\ Rev.\ D {\bf 49} (1994) 2233.
  
  \bibitem{McLerran:1993ka}
  L.~D.~McLerran and R.~Venugopalan,
  Phys.\ Rev.\ D {\bf 49} (1994) 3352.
  
  \bibitem{McLerran:1994vd}
  L.~D.~McLerran and R.~Venugopalan,
  Phys.\ Rev.\ D {\bf 50} (1994) 2225.
  
  \bibitem{Iancu:2000hn}
  E.~Iancu, A.~Leonidov and L.~D.~McLerran,
  Nucl.\ Phys.\ A {\bf 692} (2001) 583.
  
  \bibitem{Mrowczynski:2017kso}
  S.~Mrowczynski,
  Eur.\ Phys.\ J.\ A {\bf 54} (2018) no.3,  43.
  
  \bibitem{Ruggieri:2018ies}
  M.~Ruggieri and S.~K.~Das,
  EPJ Web Conf.\  {\bf 192} (2018) 00017.
  
    
  \bibitem{Song:2015jmn}
  T.~Song and T.~Epelbaum,
  arXiv:1512.05625 [nucl-th].
  
  
  \bibitem{Wong:1970fu}
  S.~K.~Wong,
  Nuovo Cim.\ A {\bf 65} (1970) 689.
  
  \bibitem{Kunihiro:2010tg}
  T.~Kunihiro, B.~Muller, A.~Ohnishi, A.~Schafer, T.~T.~Takahashi and A.~Yamamoto,
  Phys.\ Rev.\ D {\bf 82} (2010) 114015.
  f
  \bibitem{Iida:2014wea}
  H.~Iida, T.~Kunihiro, A.~Ohnishi and T.~T.~Takahashi,
  arXiv:1410.7309 [hep-ph].
  
  \bibitem{Ruggieri:2017ioa}
  M.~Ruggieri, L.~Oliva, G.~X.~Peng and V.~Greco,
  Phys.\ Rev.\ D {\bf 97} (2018) no.7,  076004.
  
  \bibitem{Lappi:2007ku}
  T.~Lappi,
  Eur.\ Phys.\ J.\ C {\bf 55} (2008) 285.
  
  \bibitem{Kowalski:2007rw}
  H.~Kowalski, T.~Lappi and R.~Venugopalan,
  Phys.\ Rev.\ Lett.\  {\bf 100} (2008) 022303.
  
  
  
  \bibitem{Lappi:2006fp} 
  T.~Lappi and L.~McLerran,
  Nucl.\ Phys.\ A {\bf 772}, 200 (2006).
  
  
  
  
  \bibitem{Moore:2004tg}
  G.~D.~Moore and D.~Teaney,
  Phys.\ Rev.\ C {\bf 71} (2005) 064904.
  
  \bibitem{Cao:2011et}
  S.~Cao and S.~A.~Bass,
  Phys.\ Rev.\ C {\bf 84} (2011) 064902.
  
  \bibitem{vanHees:2007me}
  H.~van Hees, M.~Mannarelli, V.~Greco and R.~Rapp,
  Phys.\ Rev.\ Lett.\  {\bf 100} (2008) 192301.
  
  \bibitem{Gossiaux:2008jv}
  P.~B.~Gossiaux and J.~Aichelin,
  Phys.\ Rev.\ C {\bf 78} (2008) 014904.
  
  \bibitem{Gossiaux:2009mk}
  P.~B.~Gossiaux, R.~Bierkandt and J.~Aichelin,
  Phys.\ Rev.\ C {\bf 79} (2009) 044906.
  
  \bibitem{Cao:2015hia}
  S.~Cao, G.~Y.~Qin and S.~A.~Bass,
  Phys.\ Rev.\ C {\bf 92} (2015) no.2,  024907.
  
  \bibitem{Xu:2017obm}
  Y.~Xu, J.~E.~Bernhard, S.~A.~Bass, M.~Nahrgang and S.~Cao,
  Phys.\ Rev.\ C {\bf 97} (2018) no.1,  014907.
  
\bibitem{He:2011qa}
  M.~He, R.~J.~Fries and R.~Rapp,
  Phys.\ Rev.\ C {\bf 86} (2012) 014903.
  
\bibitem{Alberico:2011zy}
  W.~M.~Alberico, A.~Beraudo, A.~De Pace, A.~Molinari, M.~Monteno, M.~Nardi and F.~Prino,
  Eur.\ Phys.\ J.\ C {\bf 71} (2011) 1666.
  
  \bibitem{Alberico:2013bza}
  W.~M.~Alberico, A.~Beraudo, A.~De Pace, A.~Molinari, M.~Monteno, M.~Nardi, F.~Prino and M.~Sitta,
  Eur.\ Phys.\ J.\ C {\bf 73} (2013) 2481.
  
  \bibitem{Ruggieri:2013ova}
  M.~Ruggieri, F.~Scardina, S.~Plumari and V.~Greco,
  Phys.\ Rev.\ C {\bf 89} (2014) no.5,  054914.
  
  \bibitem{Plumari:2015cfa}
  S.~Plumari, G.~L.~Guardo, F.~Scardina and V.~Greco,
  Phys.\ Rev.\ C {\bf 92} (2015) no.5,  054902.
  
  \bibitem{Plumari:2019gwq}
  S.~Plumari,
  Eur.\ Phys.\ J.\ C {\bf 79} (2019) no.1,  2.
  
  \bibitem{Das:2012ck}
  S.~K.~Das, V.~Chandra and J.~e.~Alam,
  J.\ Phys.\ G {\bf 41} (2013) 015102.
  
  \bibitem{Berrehrah:2013mua}
  H.~Berrehrah, E.~Bratkovskaya, W.~Cassing, P.~B.~Gossiaux, J.~Aichelin and M.~Bleicher,
  Phys.\ Rev.\ C {\bf 89} (2014) no.5,  054901.
  
  \bibitem{Berrehrah:2014kba}
  H.~Berrehrah, P.~B.~Gossiaux, J.~Aichelin, W.~Cassing and E.~Bratkovskaya,
  Phys.\ Rev.\ C {\bf 90} (2014) no.6,  064906.
  
  \bibitem{Borsanyi:2010cj}
  S.~Borsanyi, G.~Endrodi, Z.~Fodor, A.~Jakovac, S.~D.~Katz, S.~Krieg, C.~Ratti and K.~K.~Szabo,
  JHEP {\bf 1011} (2010) 077.
  
  \bibitem{Prino:2016cni}
  F.~Prino and R.~Rapp,
  J.\ Phys.\ G {\bf 43} (2016) no.9,  093002.
  
  \bibitem{Barbano:2017bcu}
  A.~Barbano [ALICE Collaboration],
  Nucl.\ Phys.\ A {\bf 967} (2017) 612.
  
  \bibitem{Acharya:2017qps}
  S.~Acharya {\it et al.} [ALICE Collaboration],
  Phys.\ Rev.\ Lett.\  {\bf 120} (2018) no.10,  102301.
  
  \bibitem{Das:2015aga}
  S.~K.~Das, M.~Ruggieri, S.~Mazumder, V.~Greco and J.~e.~Alam,
  J.\ Phys.\ G {\bf 42} (2015) no.9,  095108.
  
  
  \end{thebibliography}
\end{document}